\documentstyle[prl,aps,psfig,twocolumn]{revtex}
\newlength{\defaultparindent}
\begin{document}
\draft
\preprint{HEP/123-qed}
\title{A MEASUREMENT OF $\alpha_s (Q^2)$ FROM THE GROSS-LLEWELLYN SMITH 
SUM RULE}

\author{J.~H.~Kim$^2$, D.~A.~Harris$^6$, 
        C.~G.~Arroyo$^2$, L.~de~Barbaro$^5$, P.~de~Barbaro$^6$, 
        A.~O.~Bazarko$^2$, R.~H.~Bernstein$^3$, A.~Bodek$^6$, 
        T.~Bolton$^4$, H.~Budd$^6$, J.~Conrad$^2$, 
        R.~A.~Johnson$^1$, B.~J.~King$^2$, T.~Kinnel$^7$, 
        M.~J.~Lamm$^3$, W.~C.~Lefmann$^2$, W.~Marsh$^3$, 
        K.~S.~McFarland$^3$, C.~McNulty$^2$, S.~R.~Mishra$^2$, 
        D.~Naples$^4$, P.~Z.~Quintas$^2$, A.~Romosan$^2$, 
        W.~K.~Sakumoto$^6$, H.~Schellman$^5$, F.~J.~Sciulli$^2$,
        W.~G.~Seligman$^2$, M.~H.~Shaevitz$^2$, W.~H.~Smith$^7$,
        P.~Spentzouris$^2$, E.~G.~Stern$^2$, M.~Vakili$^1$,
        U.~K.~Yang$^6$, J.~Yu$^3$}

\address{(1)~University of Cincinnati, Cincinnati, OH 45221 USA; \\
(2)~Columbia University, New York, NY 10027 USA; \\
(3)~Fermi National Accelerator Laboratory, Batavia, IL 60510 USA; \\
(4)~Kansas State University, Manhattan, KS 66506 USA; \\
(5)~Northwestern University, Evanston, IL 60208 USA; \\
(6)~University of Rochester, Rochester, NY 14627 USA; \\
(7)~University of Wisconsin, Madison, WI 53706 USA }


\date{\today}
\maketitle

\begin{abstract}
We extract a set of values for the Gross-Llewellyn Smith sum rule at
different values of 4-momentum transfer squared ($Q^{2}$), by
combining revised CCFR neutrino data with data from other neutrino
deep-inelastic scattering experiments for $1<Q^{2}<15~~GeV^2/c^2$.  A 
comparison with the order $\alpha^{3}_{s}$ theoretical predictions
yields a determination of $\alpha_{s}$ at the scale of the Z-boson
mass of $0.114 \pm^{.009}_{.012}$~.~  This measurement provides a 
new and useful test of perturbative QCD at low $Q^2$, because of the
low uncertainties in the higher order calculations.
\end{abstract}



	The Gross-Llewellyn Smith (GLS) sum rule\cite{GLS} predicts 
the integral $\int_0^1(xF_3)\frac{dx}{x}$, where $xF_3(x,Q^2)$ is the 
non-singlet structure function measured in neutrino-nucleon ($\nu N$)
scattering.  In the naive quark parton model, the value of this
integral should be three, the number of valence quarks in the nucleon.
In perturbative Quantum Chromodynamics (pQCD), this integral is a
function of $\alpha_s(Q^2)$, the strong coupling constant.  

	The GLS integral is one of the few physical quantities which
has been calculated to next-to-next-to-leading-order (NNLO) of
perturbative QCD\cite{NNLO}, and there are estimates of the next
order term\cite{ALF4} (i.e. $O(\alpha_s ^4)$).  In addition, there is
a non-perturbative higher-twist contribution, proportional to
$1/Q^2$.  This yields the GLS integral as a function of $\alpha_s$, of
the form:  
\begin{equation}
GLS = 3 \left[ 1 - \frac{\alpha_s}{\pi} 
	- a(n_f)(\frac{\alpha_s}{\pi})^2 
	- b(n_f)(\frac{\alpha_s}{\pi})^3 \right]
	- \frac{\Delta HT}{Q^2}
 \label{eq:gls}
\end{equation}
where $a(n_f)$ and $b(n_f)$ are functions\cite{NNLO} of the number of
quark flavors accessible at a given $Q^2$.  The higher-twist
correction term $\Delta HT$ is predicted to be significant in some
models\cite{HT}, while others\cite{HT2,HT3,HT4} predict a negligibly
small correction term.  We take $\Delta HT$ as half the largest
model prediction, with errors which cover the full range ($\Delta
HT=0.15\pm .15~GeV^2$). 

	The size and $Q^2$-variation of the GLS integral is a
robust prediction in pQCD.  The NNLO calculation has been shown to be
largely independent of renormalization scheme\cite{RS}, and $xF_3$ is 
inherently independent of the gluon distribution.  The number of
orders to which the integral has been calculated ensures an accurate
perturbative calculation in spite of the large value of $\alpha_s$ at 
low $Q^2$.  

	An earlier measurement of the GLS integral has been published
by the CCFR collaboration\cite{OLDGLS}.  That analysis used a
leading-order(LO) QCD-based fit to extrapolate all data to
$Q^2=3~GeV^2$, fitted the extrapolated data to a single function over 
all $x$, and numerically integrated that function.  This was
confirmed by a LO global fit analysis of the same data\cite{JACOBI}.
However, these approaches cannot make full use of the accuracy of the 
NNLO calculation shown above, since they depend on LO pQCD for
extrapolation.  Also, the previous CCFR analysis did not correct for
quark mass thresholds\cite{RS}, target mass\cite{TGTMASS} or 
higher-twist\cite{HT,HT2,HT3,HT4} effects.  These corrections are
important at the effective mean $Q^2$ of the result 
($Q^2 \sim 3~GeV^2$). 

	This paper describes a new GLS analysis, which uses revised
CCFR $xF_3$ data together with data from earlier neutrino-scattering
experiments.  By combining data sets, we expand the kinematic region
to measure $\int xF_3\frac{dx}{x}$ for $1<Q^2<15~GeV^2$ without any
extrapolation in $Q^2$.  This technique thus allows us to consistently
use the fundamental NNLO prediction shown in equation (1).  

	The CCFR data were collected at Fermilab in experiments E744
and E770, which ran in 1985 and 1987-8 respectively.  The experiments
observed neutrino scattering in an iron calorimeter\cite{DETECT1}.
The calorimeter and muon spectrometer were calibrated using a test
beam\cite{DETECT2}.  New structure functions (SFs) from this 
data\cite{NEWSF} were published in 1997.  These SFs had a number
of improvements compared to the SFs used in the previous GLS
measurement\cite{OLDGLS}.  Improvements include a revised energy
calibration based directly on test-beam data, an improved calculation
of radiative corrections\cite{BARDIN}, and the removal of and
correction for two-muon events ($\nu N \rightarrow \mu^+ \mu^- X$)
from the data sample.  Previously, the two-muon events introduced a
small ambiguity at low $x$ between neutrino-induced and
anti-neutrino-induced events, which is particularly important to the
GLS integral.  
 
	This analysis further improves the CCFR structure
functions\cite{NEWSF} at low $x$ by improving the acceptance and
smearing corrections.  These corrections, which require a
cross-section model, now incorporate measurements of the strange
sea\cite{DIMUON} and a more accurate parameterization of the parton
distributions.  These procedures create a new SF set\cite{GLSSF} with
reduced uncertainty at low $x$. 

	We also expand our kinematic region at high $x$ by using 
$xF_3$ data from other $\nu$-N experiments, namely: WA25\cite{WA25},
WA59\cite{WA59}, SKAT\cite{SKAT}, FNAL-E180\cite{E180}, and
BEBC-Gargamelle\cite{BEBC}.  These were each normalized to CCFR in the
regions of overlap, and the WA25 data were corrected at high $x$ for
nuclear differences\cite{EMC} in the targets.  

	The GLS integral is evaluated numerically using the combined 
$xF_3$ data in bins of $x$ and $Q^2$.  The integral over $x$ is 
evaluated separately for each $Q^2$ bin.  At very low $x$ we must
extrapolate below the CCFR kinematic limit, while at high $x$ we use
other experiments' data, and interpolate as necessary within the large
bins.  In each case, we vary the forms of the interpolations and
extrapolations, and use the differences in the integral as estimates
of the systematic uncertainties in the procedures.  

	The CCFR data have a minimum $x$ of roughly ($x=0.002 \times
Q^2$).  To extrapolate below this, we fit a power law ($Ax^B$) to
all points with $x<0.1$.  The power-law form is suggested by Regge
theory\cite{REGGE}, which predicts a shape of $x^{0.5}$.  To test
this assumption, we made an alternate fit of the form $Cx^{0.5}$,
using the difference as an independent systematic uncertainty.  This 
systematic error becomes large at $Q^2$ above $5~GeV^2$.  

	For $x>0.5$, there are also few data points.  Most of the 
data here come from BEBC and SKAT which quote only two points for the
range $0.5<x<1.0$.  Here $xF_3$ is steeply falling and thus the
precise shape is important for integration.  Again, to estimate the
contribution and error we use various assumptions.  For the central
value, we use the principle that at high $x$, the shape of $xF_3$
should be the same as $F_2$, since the sea quarks are negligible at
high $x$.  Electron scattering experiments at SLAC have precisely
measured $F_2$ in this region\cite{SLAC}.  These data are corrected
for nuclear effects\cite{EMC} and differences between $eN$ and $\nu N$ 
scattering\cite{NEWSF}.  The corrected $F_2$ data by itself give 
the same result as interpolating the $xF_3$ data with the $F_2$
shape.  

%
%
	However, the SLAC data have small resonance peaks which may be
different from neutrino resonances.  To estimate the systematic error, 
we take the difference between two power-law fits to the $xF_3$ data,
using the forms $D(1-x)^E$ and $F(1-x)^3$.  These bracket the SLAC fit
and serve as the limits of reasonable interpolation forms. Note that
resonance behavior at low $Q^2$ coupled with approximate
scaling\cite{QUARK} leads to a predicted form of $(1-x)^3$. 

\begin{figure}
\centerline{\psfig{figure=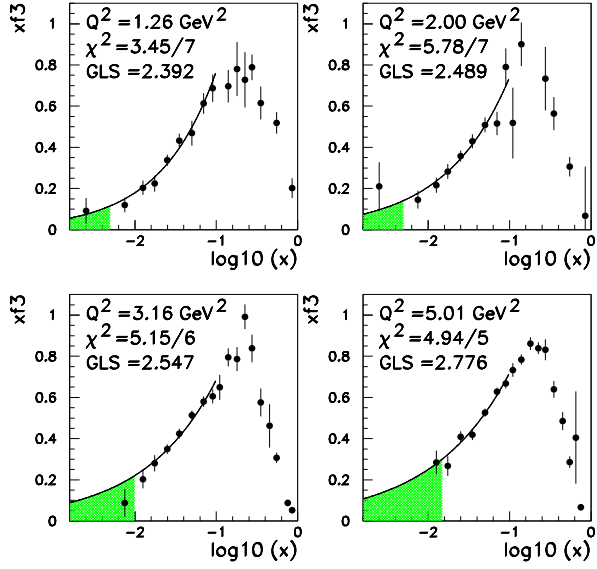,width=3.375in}}
\caption{$xF_3$ as a function of $x$ at the four lowest $Q^2$ values, 
with $x$ on a log scale.  The area under the points thus represents 
$\int xF_3\frac{dx}{x}$.  The curve is a power law($Ax^B$) fit to the
$x<0.1$ points, which is used to calculate the integral in the shaded
region.}
\label{fig1}
\end{figure}

\begin{figure}
\centerline{\psfig{figure=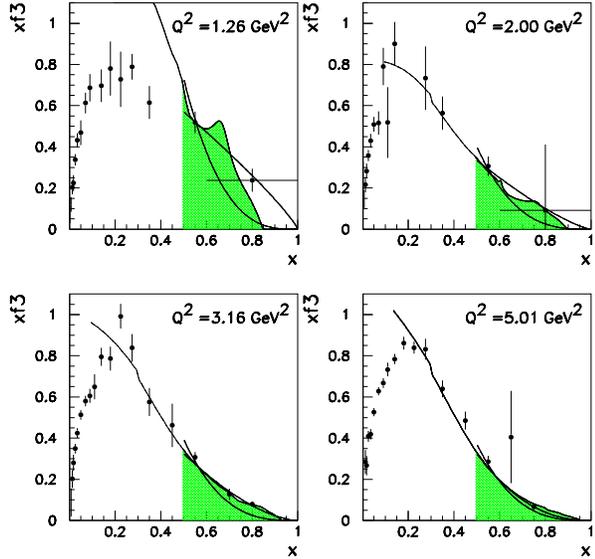,width=3.375in}}
\caption{$xF_3$ as a function of $x$ at the four lowest $Q^2$ values, 
with $x$ on a linear scale to show the high-$x$ data.  The shaded
region shows the fit using the shape from SLAC $F_2$ data.  The other 
lines are the power law fits ($D(1-x)^E$ above and $F(1-x)^3$ below)
used to estimate systematic error.}
\label{fig2}
\end{figure}

	Fig.~\ref{fig1} shows the combined $xF_3$ data on a log $x$
scale in four low-$Q^2$ regions, along with a line representing the
power law fit ($Ax^B$) for $x<0.1$ and the $\chi^2$ for the fits.
Fig.~\ref{fig2} shows the same data on a linear $x$ scale to
highlight the high-$x$ region.  

	Following the procedure of the BEBC collaboration\cite{BEBC},
we add the quasi-elastic contribution and correct the GLS integral for
target mass effects.  This is necessary to be consistent with
theoretical prediction of higher-twist contributions\cite{HT} to the
GLS integral.  Table~\ref{table1} shows the exact $Q^2$ ranges of each
bin and the contributions to $\int xF_3\frac{dx}{x}$ for the different 
regions of $x$.  

	The systematic uncertainties are divided into three classes.
The first includes calibration, normalization, and other purely
experimental issues.  The second is uncertainty in the integration of
experimental $xF_3$, estimated by varying the assumed functional forms 
as described above.  The third class includes uncertainties in the
theoretical prediction of the GLS integral itself.  A summary of all
the uncertainties is shown in Table~\ref{table2}. 

	The dominant experimental systematic uncertainties are in the
normalization of $xF_3$, which comes from the total neutrino and
anti-neutrino cross-sections ($\sigma _{\nu}$ and $\sigma _{\bar
\nu}$). The absolute $\sigma _{\nu}$ is not measured by CCFR, so we
use the world average\cite{XSEC,XSEC2}
($\sigma _{\nu}/E_{\nu} = 0.677 \pm .007 \times 10^{-38} cm^2/GeV$). 
The ratio $\sigma _{\bar \nu} / \sigma _{\nu}$ is measured by CCFR,
and combined with the world average\cite{XSEC2,XSECRAT} yields 
$\sigma _{\bar \nu} / \sigma _{\nu} = 0.499 \pm .007$.   
Other experimental uncertainties include the energy scale calibration
of the detector and the effects of charm production on the measured
structure functions.  

	Additionally, there is a small uncertainty in the revised
calculation of acceptance and smearing corrections.  These corrections
depend on a parameterization of the SFs.  Variations of the functional
form of the parameterization were used to estimate the systematic
error.  

	The dominant theoretical uncertainty is the error on the
higher-twist correction ($\Delta HT/Q^2$).  Braun and
Kolesnichenko\cite{HT} use three models which predict a correction
term $\Delta HT$ between $0.16$ and $0.29~GeV^2$.  Other models, such
as bag models\cite{HT2} and a recent NNLO analysis\cite{HT3} using a
renormalon\cite{HT4} approach, predict a negligible correction term  
($\Delta HT<0.02~GeV^2$).  For our central value, we take 
$\Delta HT=0.15\pm .15~GeV^2$, thus covering all three predictions.
The nuclear effects of the target are predicted to be
small\cite{HTNUC} ($-0.01/Q^2$ for iron).  We also use estimates of
the renormalization scheme dependence\cite{RS} and the order $\alpha_s
^4$ term in the pQCD expansion\cite{ALF4} as uncertainties in the
perturbative calculation.  

	To extract a single value for $\Lambda_{\overline{MS}}^{(5)}$,
we combine the measured values of the GLS integral in each $Q^2$ with 
the uncorrelated systematic errors, including the acceptance model
error and the high-$x$ and low-$x$ fitting errors.  These points are
fitted to the NNLO pQCD function and higher-twist term, shown in
Eq.~(\ref{eq:gls}).  The prediction includes quark mass thresholds 
using the procedure of Chyla and Kataev\cite{RS}.  The other
systematic error sources are fully correlated in $Q^2$, and are
applied by shifting all GLS values by the uncertainty from that 
source and redoing the fit.  The difference between the shifted 
and unshifted fit result represents the uncertainty in $\alpha_s$ 
from that systematic error.  

	The best fit to the measured GLS integral as a function of
$Q^2$ is for a value of $\Lambda_{\overline{MS}}^{(5)}=165~MeV$.
Evolving to $M_Z^2$ and $3\,GeV^2$ at NNLO, this corresponds to: 
\begin{eqnarray}
\alpha_s(M_Z^2) & = & 0.114 \pm^{.005}_{.006}(stat)
                      \pm^{.007}_{.009}(syst) 
                      \pm.005(thry)
\end{eqnarray}
\begin{eqnarray}
\alpha_s(3~GeV^2) & = & 0.28 \pm.035 (stat)
                      \pm.05 (syst)
                      \pm^{.035}_{.03}(thry)
\end{eqnarray}
If the higher twist models of [5-7] are used, the central values 
become $\alpha_s(M_Z^2)=0.118$ and $\alpha_s(3\,GeV^2)=0.31$.  
Table III shows the results of fitting for
$\Lambda_{\overline{MS}}^{(5)}$ at each Q2 value as a consistency
check.  In all cases the small target mass and quasi-elastic
corrections are included, and roughly cancel.  

	In conclusion, the GLS sum rule allows a precise measurement
of $\alpha_s$ at low $Q^2$.  An independent measurement of $\alpha_s$
from the CCFR calculation\cite{NEWSF} used the slope of global NLO
fits to $xF_3$ and $F_2$ for $15<Q^2<125~GeV^2$, and found 
$\alpha_s(M_Z^2)=0.119\pm .004$.  Three orders of magnitude higher in
scale ($Q^2=M_Z^2=8315~GeV^2$), electroweak fits to LEP data\cite{LEP}
found $\alpha_s(M_Z^2)=0.124\pm .004 \pm .002$ based on the parameter 
$R_{\ell}$.  

	The GLS result is consistent with other measurements, showing
the power of pQCD across a very wide range of scales.  An
inconsistency in results might have indicated a need for higher order 
calculations in other measurements (i.e. NNLO) or the presence of
other theoretical effects\cite{SUPER} which scale with $Q^2$.  The GLS 
result can be improved in the future by additional data from the NuTeV
experiment at Fermilab and by better understanding of higher-twist
effects.  

\acknowledgments

\noindent
{\it Acknowledgments.}  This research was supported by the
U.S. Department of Energy and the National Science Foundation.  We 
thank the staff of the Fermi National Accelerator Laboratory for
their hard work in support of this experimental effort.  We also wish
to thank Andrei Kataev for helpful suggestions regarding the
theoretical calculations. 


\newpage

\newpage
\renewcommand{\arraystretch}{1.1}  

\begin{table}
\caption{The contributions to the GLS integral from different regions
of $x$ and the quasi-elastic peak added at $x=1$, shown as a function
of $Q^2$ (in $GeV^2$).  For high and low $x$, it shows the estimated
uncertainties due to the model choice only.  The values include the
target mass corrections.}

\begin{tabular}{c|c|c|c|l}
\rule[-1.5ex]{0ex}{4ex} $Q^2$ & $\int_{0}^{.02}F_3dx$ & $\int_{.02}^{.5}F_3dx$ & $\int_{.5}^{1.}F_3dx$ & qElas \\ \hline \hline
~1.0-~1.6 & $0.376\pm.082$ & 1.730 & $0.183\pm.073$ & 0.103 \\
~1.6-~2.5 & $0.523\pm.002$ & 1.843 & $0.091\pm.026$ & 0.033 \\
~2.5-~4.0 & $0.558\pm.026$ & 1.889 & $0.092\pm.020$ & 0.009 \\
~4.0-~6.3 & $0.700\pm.137$ & 1.991 & $0.084\pm.016$ & 0.002 \\
~6.3-10.0 & $0.748\pm.139$ & 2.004 & $0.064\pm.008$ & 0.0004 \\
10.0-15.5 & $0.718\pm.113$ & 2.007 & $0.073\pm.003$ & 0.0001 \\
\end{tabular}
\label{table1}
\end{table}
%
%
\renewcommand{\arraystretch}{1.3}  

\begin{table}
\caption{Uncertainties in $\alpha_s(M_Z^2)$.  Errors which are
uncorrelated in $Q^2$ are marked with a $^*$.  Other sources are 
fully correlated in $Q^2$.}

\begin{tabular}{|l|c|}
{\bf Source}			& {\bf Error} \\ \hline
{\bf Statistical} 		& $\left(~^{+.005}_{-.006}\right)$  \\ \hline
$\sigma_{\nu}$ Normalization	& $\left(~^{+.003}_{-.005}\right)$  \\
$\sigma_{\nu}/\sigma_{\bar \nu}$ Ratio & $\left(~^{+.005}_{-.006}\right)$ \\
Energy Calibration		& $\left(~^{+.002}_{-.003}\right)$  \\
Charm Production		& $\pm.0005$  \\
Acceptance model$^*$ 		& $\pm.002$  \\ \hline
{\bf Total Experimental Error} 	& $\left(~^{+.006}_{-.008}\right)$  \\ \hline
High-$x$ fitting$^*$		& $\pm.003$  \\
Low-$x$ fitting$^*$		& $\pm.002$  \\ \hline
{\bf Total Model Error}		& $\pm.004$  \\ \hline
{\bf Combined Systematic Error}	& $\left(~^{+.007}_{-.009}\right)$  \\ \hline
Higher-twist			& $\left(~^{+.004}_{-.005}\right)$  \\
Renormalization Scheme\cite{RS}	& $\pm.001$ \\
Order $\alpha_s^4$ 		& $\pm.0003$  \\ \hline
{\bf Total Theory Error} 	& $\pm.005$ \\
\end{tabular}
\label{table2}
\end{table}
%
%

\begin{table}
\caption{The total GLS integral and $\alpha_s$ for each bin in
$Q^2$. The errors on the GLS are $\pm(stat)\pm(syst)$.  The errors
on $\alpha_s(Q^2)$ are $\pm(stat)\pm(syst)\pm(thry)$.  Systematic
errors are correlated in $Q^2$. }

\begin{tabular}{c|c|c}
$\langle Q^2 \rangle$ & $GLS(Q^2)$ & $\alpha_s(Q^2)$ \\ \hline
 1.26 & $2.39\pm.08 \pm.14$ & $0.330\pm.023 \pm.042 \pm.050$ \\
 2.00 & $2.49\pm.08 \pm.10$ & $0.303\pm.020 \pm.026 \pm.036$ \\
 3.16 & $2.55\pm.06 \pm.10$ & $0.287\pm.008 \pm.034 \pm.026$ \\
 5.01 & $2.78\pm.06 \pm.19$ & $0.165\pm.033 \pm.144 \pm.024$ \\
 7.94 & $2.82\pm.07 \pm.19$ & $0.145\pm.061 \pm.136 \pm.022$ \\
12.59 & $2.80\pm.13 \pm.18$ & $0.164\pm.068 \pm.101 \pm.014$ \\
\end{tabular}
\label{table3}
\end{table}

\end{document}